\documentclass[onecolumn]{article}

\usepackage{amsmath,amssymb,amsfonts,amsthm,enumerate,multirow,mathtools}
\usepackage{color}
\usepackage{siunitx}
\usepackage{parskip}
\usepackage[hidelinks]{hyperref}
\usepackage{placeins} 
\usepackage{booktabs}
\usepackage{algorithm,algpseudocode}
\usepackage[affil-it]{authblk}
\usepackage[margin=2.5cm]{geometry}
\usepackage[round]{natbib}

\newcommand{\pmat}[1]{\begin{bmatrix}#1\end{bmatrix}}
\newcommand{\gd}[3]{\frac{\sin(\alpha_{d_0 x} #1 \alpha_{\varphi x} #2 \alpha_{d_0 y}#3 \alpha_{\varphi y}) }{8(\hat{n}_1\lambda_{d_0 x,k} #1 \hat{n}_1\lambda_{\varphi x,j} #2 \hat{n}_2\lambda_{d_0 y,k} #3 \hat{n}_2\lambda_{\varphi y,j})}}
\newcommand{\Transp}{\mathsf{T}}

\title{Neutron Transmission Strain Tomography for Non-Constant Stress-Free Lattice Spacing}

\author[1]{J.N. Hendriks}
\author[2]{C. Jidling}
\author[2]{T.B. Sch{\"o}n}
\author[1]{A. Wills}
\author[1]{C.M. Wensrich}
\author[1]{E.H. Kisi}

\affil[1]{School of Engineering, The University of Newcastle, Callaghan NSW 2308, Australia}
\affil[2]{Department of Information Technology, Uppsala University, Sweden}

\begin{document}
\newcommand{\coverTitle}{Neutron Transmission Strain Tomography for Non-Constant Stress-Free Lattice Spacing}
\newcommand{\coverAuthors}{J.N. Hendriks, C. Jidling, T.B. Sch{\"o}n, A. Wills, C.M. Wensrich, E.H. Kisi}
\newcommand{\coverStatus}{Accepted for publication.}

\begin{titlepage}
    \begin{center}
        {\large \em Technical report}
        
        \vspace*{2.5cm}
        %
        {\Huge \bfseries \coverTitle  \\[0.4cm]}
        
        %
        {\Large \coverAuthors \\[2cm]}
        
        \renewcommand\labelitemi{\color{red}\large$\bullet$}
        \begin{itemize}
            \item {\Large \textbf{Please cite this version:}} \\[0.4cm]
            \large
            \coverAuthors. \coverTitle. \textit{Nuclear instruments and methods in physics research section B}, 456:64-73, 2019.  
        \end{itemize}
        
        \vfill

        
        \vfill
    \end{center}
\end{titlepage}

\maketitle

\begin{abstract}
Recently, several algorithms for strain tomography from energy-resolved neutron transmission measurements have been proposed.
These methods assume that the stress-free lattice spacing $d_0$ is a known constant limiting their application to the study of stresses generated by manufacturing and loading methods that do not alter this parameter.
In this paper, we consider the more general problem of jointly reconstructing the strain and $d_0$ fields. 
A method for solving this inherently non-linear problem is presented that ensures the estimated strain field satisfies equilibrium and can include knowledge of boundary conditions.
This method is tested on a simulated data set with realistic noise levels, demonstrating that it is possible to jointly reconstruct $d_0$ and the strain field.

\end{abstract}



\section{Introduction} 
\label{sec:background}
Energy-resolved neutron transmission methods can generate lower dimensional (one- or two-dimensional) images of strain from a higher dimensional (two- or three-dimensional) strain field within a polycrystalline material. 
The `tomographic' reconstruction of an unknown strain field from these images can be used to study the residual strain and stress within engineering components.
Residual stresses are those which remain after applied loads are removed (e.g. due to heat treatment, plastic deformation, etc.), and may have significant and unintended impact on a component's effective strength and service life --- in particular its fatigue life.
Measuring and quantifying these strains is important for the validation of predictive design tools, such as Finite Element Analysis, and to aid the development of novel manufacturing techniques --- i.e. additive manufacturing.

These strain images are generated by analysing features known as Bragg-edges in the relative transmission of a neutron pulse through a sample. 
Bragg-edges are sudden increases in the intensity as a function of wavelength and occur when the scattering angle $2\vartheta$ reaches $180^\circ$, beyond which no further coherent scattering can occur.
The wavelength $\lambda$ at which these Bragg-edges occur can be related to the lattice spacing $d$ within the sample through Bragg's law: $\lambda = 2d\sin\vartheta$.
Assuming minimal texture, this can be used to provide a relative measure of strain;
\begin{equation}\label{eq:relative_strain}
    \langle \epsilon \rangle = \frac{d-d_0}{d_0},
\end{equation}
where $d_0$ is the stress-free lattice spacing and $\langle\epsilon \rangle$ is a through thickness average of the normal, elastic strain in the direction of the beam.

The determination of $d_0$ is a problem inherent to diffraction and transmission strain analysis.
For specific cases where the loading mechanism does not result in changes to the stress-free lattice parameter, its value can be measured prior to loading and in the simplest case (e.g. for an annealed sample) a constant value throughout the sample can be assumed.
Several algorithms for strain tomography assuming a known, constant stress-free lattice spacing have been developed. 
Reconstruction of axisymmetric strain fields is considered in \citep{abbey09,abbey12,kirkwood15,gregg2017tomographic} and more general two-dimensional strain fields in \citep{gregg2018tomographic,jidling2018probabilistic,hendriks2018traction}. 

Many manufacturing techniques (e.g. welding and additive manufacturing) can alter the lattice spacing; for example, as a result of inhomogeneously distributed phase changes (such as the Martensite transformation), or due to gradients in composition as a result of differing chemical states in the starting materials.
Since the lattice spacing (in this case $d_0$) is sensitive to crystal structure and composition changes, the stress-free lattice parameter may vary throughout the sample. 
Ignoring variations in $d_0$ would cause severe degradation in the quality of a reconstructed strain field.
In such cases, measuring $d_0$ is more challenging and has been achieved in neutron diffraction measurements by measuring additional directions of strain
\citep{luzin2011residual,choi2007integrated} and by destructive methods where the strain is relieved by wire cutting the sample into a grid allowing the stress-free lattice spacing to be measured throughout the sample \citep{paradowska2005neutron}.
Although the latter of these two options could be applied to strain tomography it requires the destruction of the sample and creates an additional tomography problem, requiring another set of measurements to be acquired.

Here, we present a method capable of jointly reconstructing the strain field and the $d_0$ field from a single set of neutron transmission images. 
To achieve this both the strain and $d_0$ are modelled by a Gaussian process (see for example \citet{rasmussen2006gaussian}) and equilibrium and boundary conditions are built into the strain model \citep{JidlingWWS:2017}.
This extends the Gaussian process approach presented by \citet{jidling2018probabilistic,hendriks2018traction} to handle the inherently non-linear nature of this problem.
A numerically tractable algorithm based on variational inference (see for example \citet{BleiKM:2017,jordan1999introduction}) is provided and the method is validated on a simulated data set.

\section{Problem Statement} 
\label{sec:problem_statement}
This paper focuses on the joint reconstruction of the strain field $\boldsymbol\epsilon(\mathbf{x})$ and a non-constant stress-free lattice parameter $d_0(\mathbf{x})$ from a set of neutron transmission images.
Restricting the problem to two dimensions, gives the strain field as the symmetric tensor
\begin{equation}
    \boldsymbol\epsilon(\mathbf{x}) = \pmat{\epsilon_{xx}(\mathbf{x}) & \epsilon_{xy}(\mathbf{x}) \\ \epsilon_{xy}(\mathbf{x}) & \epsilon_{yy}(\mathbf{x})},
\end{equation}
where $\mathbf{x} = \pmat{x & y}^\Transp$. For brevity, the unique components of strain will be written as $\bar{\epsilon} = \pmat{\epsilon_{xx} & \epsilon_{xy} & \epsilon_{yy}}^\Transp$ with the coordinate $\mathbf{x}$ omitted where appropriate.

\begin{figure}[!ht]
    \centering
    \includegraphics[width=0.3\linewidth]{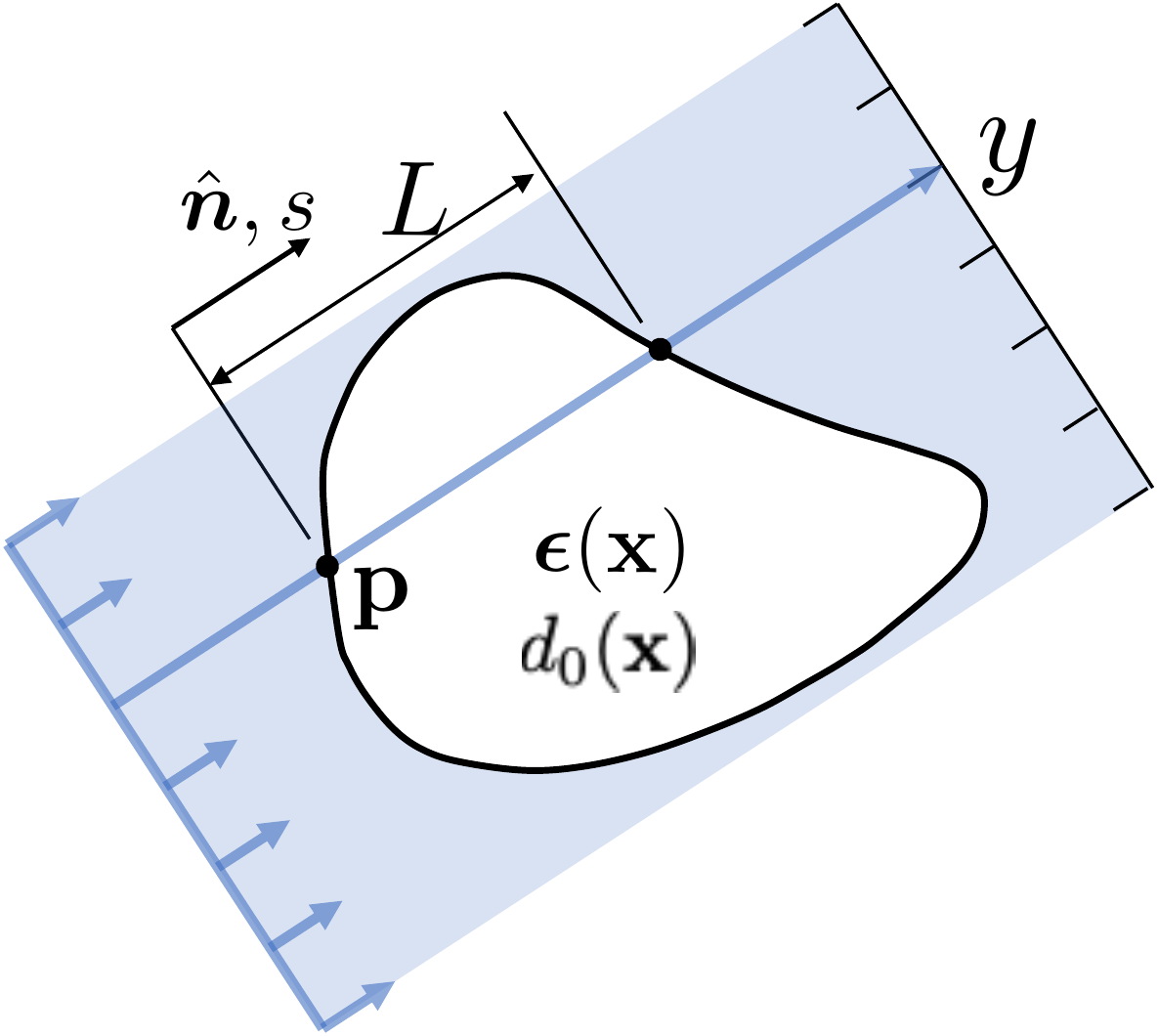}
    \caption{LRT measurement geometry. Each measurement made by a detector pixel is associated with a ray of direction $\hat{\mathbf{n}}$ that enters the sample at $\mathbf{p}$ and has a total irradiated length of $L$.}
    \label{fig:LRT}
\end{figure}

Here, we consider the lattice spacings $d$ as the measurements rather than the standard approach which considers the relative strain of the form form \eqref{eq:relative_strain}.
This allows the measurements to be explicitly related to both the strain and the stress-free lattice parameter through the Longitudinal Ray Transform (LRT) \citep{lionheart15}:
\begin{equation}\label{eq:LRT}
    y(\boldsymbol\eta) = d(\boldsymbol\eta)+e = \frac{1}{L}\int\limits_0^L\bar{\mathbf{n}}\bar{\boldsymbol\epsilon}(\mathbf{p}+\hat{\mathbf{n}}s) d_0(\mathbf{p}+\hat{\mathbf{n}}s) + d_0(\mathbf{p}+\hat{\mathbf{n}}s) \, \mathrm{d}s + e.
\end{equation}
where $e\sim\mathcal{N}(0,\sigma_n^2)$ and the geometry of each measurement is given by the parameter set $\boldsymbol\eta=\{\hat{\mathbf{n}},L,\mathbf{p}\}$; with $\hat{\mathbf{n}}$ as the beam direction, $L$ as the irradiation length, $\mathbf{p} = \pmat{x_0 & y_0}^\Transp$ as the point of initial intersection between the ray and the sample, and $\bar{\mathbf{n}} = \pmat{\hat{n}_1^2 & 2\hat{n}_1\hat{n}_2 & \hat{n}_2^2}$. 
See Figure~\ref{fig:LRT} for the measurement geometry.
These measurements are a non-linear function of the two phenomena we wish to estimate; $\boldsymbol\epsilon$ and $d_0$.

For details on the analysis of neutron transmission data to determine these lattice spacings the reader is referred to \citet{santisteban02,santisteban02b,tremsin11,tremsin12}. It is also worth noting that the standard deviation $\sigma_n$ of these measurements is available. 

Furthermore, the strain field inside a sample is a physical property and as such it is subject to equilibrium and boundary conditions \citep{sadd2009elasticity}.
Therefore, it is natural to constrain estimates of the strain field to satisfy these conditions.
Using Hooke's law the equilibrium conditions can be written directly in terms of strain. In two dimensions, this relies on an assumption of plane strain or plane stress. Plane stress is assumed for the remainder of this work, giving the equilibrium conditions as
\begin{equation}
\begin{split}
    \frac{\partial}{\partial x}(\epsilon_{xx} + \nu\epsilon_{yy}) + \frac{\partial}{\partial y}(1-\nu)\epsilon_{xy} &= 0, \\
    \frac{\partial}{\partial x}(\epsilon_{yy} + \nu\epsilon_{xx}) + \frac{\partial}{\partial y}(1-\nu)\epsilon_{xy} &= 0, \\
\end{split}
\end{equation}
where $\nu$ is Poisson's ratio.

Boundary conditions, in particular the load free surfaces, may also be known.
For an unloaded surface, the distribution of forces known as tractions will be zero. 
Through equilibrium this places additional linear constraints on the strain field, which, assuming plane stress, can be written as
\begin{equation}\label{eq:boundary_cons}
    \mathbf{0} = \pmat{n_{\perp1} & n_{\perp2} & 0 \\ 0 & n_{\perp1} & n_{\perp2} }
    \pmat{1 & 0 & -\nu \\ 0 &1+\nu&0 \\ -\nu & 0 & 1}
    \bar{\boldsymbol\epsilon}(\mathbf{x}_b),
\end{equation}
where $\mathbf{x}_b$ is a point on an unloaded surface and $\mathbf{n}_\perp$ is the normal to the surface at this point.

An approach to enforcing equilibrium in the estimated strain field is to define a Gaussian process for the Airys stress function from which strain can be derived \citep{jidling2018probabilistic}.
This non-parametric approach was demonstrated experimentally by \citet{jidling2018probabilistic} and compared to other parametric approaches by \citet{hendriks2018traction} with promising results. 
Boundary conditions in the form of \eqref{eq:boundary_cons} can be included in the estimation process as artificial measurements of zero traction \citep{hendriks2018traction}.

We wish to extend this approach so that both the strain field and the stress-free lattice spacing can be estimated. 
As the measurements are a non-linear function of the unknowns we cannot directly apply the standard Gaussian process regression methods \citep{rasmussen2006gaussian}.
There exists several approaches to approximate Gaussian processes for non-linear functions; the Laplace approximation \citep{rasmussen2006gaussian,bishop1995neural}, GP variational inference \citep{steinberg2014extended}, and Markov Chain Monte Carlo methods (such as Elliptical Slice Sampling \citep{murray2010slice}). 
For these methods, the measurements are modelled as non-linear functions of the GP sampled at the measurement locations (known as latent function values). The latent function values that best\footnote{For a given criterion of best fit, whether it be marginal log likelihood, cross-validation, etc.} match the data are determined by one of the above methods. Then, Gaussian process regression is applied with the latent function values taking the place of measurements to determine the function values at the new locations of interest. 

The non-static nature of the integral measurement model \eqref{eq:LRT} makes it unclear how to express the measurements as a function of a finite set of latent function values, and hence the above approaches to approximating the GP for non-linear measurements cannot be applied directly. 
In the following section, we utilise an finite basis function approximation to the GP, and by viewing the problem from an alternate perspective we show how variational inference can be used to solve this non-linear problem.


\section{Method} 
\label{sec:method}
The method presented here is to define a Gaussian process model for the strain field and the stress-free lattice spacing. This Gaussian process model is then approximated using a Hilbert space approximation \citep{solin2014hilbert,jidling2018probabilistic}. 
This has two benefits; firstly it removes the need for numerical integration of the covariance function (as discussed by \citet{jidling2018probabilistic,JidlingWWS:2017}), and secondly it allows us to reformulate the problem as a set of basis functions with unknown coefficients.
Variational inference can then be used to learn the coefficients from the LRT measurements and artificial measurements of zero traction.

\subsection{Gaussian Process model} 
\label{sub:gaussian_process_prior}
The Gaussian process (GP) is a Gaussian distribution of spatially correlated functions;
\begin{equation}
    f(\mathbf{x}) \sim \mathcal{GP}\left(m(\mathbf{x}),k(\mathbf{x},\mathbf{x}')\right).
\end{equation}
The characteristics of the functions belonging to this distribution are governed by a mean function $m(\mathbf{x})$ and a covariance function $k(\mathbf{x},\mathbf{x}')$.
The covariance function describes the correlation between the function values $f(\mathbf{x})$ and $f(\mathbf{x}')$ at any two points $\mathbf{x}$ and $\mathbf{x}'$.
Careful design of the covariance function can ensure that only functions satisfying desired characteristics belong to the distribution.

Here, we wish to design the covariance function to ensure that only strain fields satisfying equilibrium are contained in the GP prior.
Following the formulation in \citep{jidling2018probabilistic} a GP model for the Airys stress functions is defined; $\varphi(\mathbf{x}) \sim\mathcal{GP}\left(0,k_\varphi(\mathbf{x},\mathbf{x}')\right)$. Under the assumption that the sample is plane stress, isotropic, and contiguous, the Airy's stress functions can be related to strain through the mapping 
\begin{equation}\label{eq:airys_to_strain}
    \bar{\boldsymbol{\epsilon}}(\mathbf{x}) = \mathcal{V}^\mathbf{x}\varphi(\mathbf{x}),
    \qquad
    \mathcal{V}^\mathbf{x} = \begin{bmatrix}
        \frac{\partial^2}{\partial y^2} - \nu\frac{\partial^2}{\partial x^2} \\
        -(1 + \nu)\frac{\partial^2}{\partial x \partial y} \\
        \frac{\partial^2}{\partial x^2} - \nu\frac{\partial^2}{\partial y^2}
    \end{bmatrix},
\end{equation}
where $\mathcal{V}$ is a linear operator, and the superscript denotes which variable the operator acts on. As GPs are closed under linear operators \citep{papoulis2002probability,hennig2013quasi,wahlstrom2015modeling,jidling2018probabilistic} a GP model for strain that satisfies equilibrium can now be defined;
\begin{equation}
    \bar{\boldsymbol\epsilon}(\mathbf{x}) \sim\mathcal{GP}\left(0,\mathcal{V}^\mathbf{x}k_\varphi(\mathbf{x},\mathbf{x}')\mathcal{V}^{\mathbf{x}'}{}^\Transp\right),
\end{equation}
where a prior mean function of zero has been chosen.

Additionally, $d_0$ function is also modelled by a GP;
\begin{equation}
    d_0(\mathbf{x}) \sim \mathcal{GP}(\bar{d}_0,k_{d_0}(\mathbf{x},\mathbf{x}')).
\end{equation}
where the prior mean $\bar{d}_0$ is chosen to be close to the expected theoretical stress-free lattice spacing for the material used or a measured average in a stress-free sample.
The choice of prior mean function does not mean that we believe the $d_0$ and $\bar{\boldsymbol{\epsilon}}$ functions to be a particular value, but rather that we do not have any information to suggest otherwise. After the inclusion of measurement information, the mean of the posterior estimate will be updated.

There exists a number of options for the base covariance functions $k_\varphi(\mathbf{x},\mathbf{x}')$ and $k_{d_0}(\mathbf{x},\mathbf{x}')$, with both the squared-exponential and the Mat{\'e}rn covariance functions having been successfully used for strain estimation \citep{jidling2018probabilistic,hendriks2018traction}. 
For a more thorough discourse on available covariance functions the reader is referred to \citet{rasmussen2006gaussian}.

Having defined suitable GP models for the strain and $d_0$ fields we now wish to estimate these fields from the LRT and traction measurements. 
However, the LRT is a non-linear function of these fields and consequently a closed form solution does not exist.
The following presents a method for obtaining these estimates that approximates the GP by a finite number of basis functions allowing variational inference to be applied.




\subsection{Hilbert Space Approximation to the GP Prior} 
\label{sub:hilbert_space_approximation_to_the_gp_prior}
Here, we make use of the approximation method proposed by \citep{solin2014hilbert} and demonstrated to be suitable for the problem of strain tomography \citep{jidling2018probabilistic}. This method approximates our covariance function by a finite sum of $m$ basis functions;
\begin{equation}
    k(\mathbf{x},\mathbf{x}') = \sum_{j=1}^{m} \phi_i(\mathbf{x}) S(\boldsymbol\lambda_j)\phi_j(\mathbf{x}'), \qquad 
\end{equation}
where $S$ is the spectral density of the covariance function. For a stationary covariance function $k = k(\mathbf{r})$, where $\mathbf{r} = \mathbf{x}-\mathbf{x}'$, the spectral density and the basis functions are given by;
\begin{equation}
    S(\boldsymbol\omega) = \int k(\mathbf{r})e^{-i\boldsymbol\omega^\Transp\mathbf{r}}\,\mathrm{d}\mathbf{r}, \quad 
    \phi_{j} = \frac{1}{\sqrt{L_{x}L_{y}}}\sin(\lambda_{x,j}(x + L_{x}))\sin(\lambda_{y,j}(y + L_{y})),
\end{equation}
where $L_x$ and $L_y$ control the domain size, and $\boldsymbol\lambda = [\lambda_x, \lambda_y]^\Transp$ encodes spatial frequencies of the basis functions. 
The basis functions are chosen as a solution to the Dirichlet boundary conditions on a rectangular domain, which is a natural choice for the Laplace eigenvalue problem that needs to be solved to approximate the GP \citep{jidling2018probabilistic}.
The parameters $\theta = \{l_x,l_y,\sigma_f\}$ are commonly called `hyperparameters' and can be chosen by optimisation (as discussed in Section~\ref{sub:hyperparameter_optimisation}).
For our application the domain size and spatial frequencies are chosen such that the basis functions spanned a region where their spectral densities, were greater than a minimum threshold. 
This helps to ensure that the dominant frequencies of the response are captured while maintaining numerical stability.

At this stage, the alternative view point of Bayesian linear regression can be taken. 
This approach models the unknown function by a set of basis functions with Gaussian coefficients;
\begin{equation}
    f(\mathbf{x}) = \sum_{j=1}^m \phi_j(\mathbf{x}) w_j = \boldsymbol\phi(\mathbf{x})\mathbf{w}, \qquad w_j \sim\mathcal{N}(\mu_j,S(\boldsymbol\lambda_j)),
\end{equation}
where $\boldsymbol{\phi}(\mathbf{x})$ and $\mathbf{w}$ have dimensions $[1,m]$ and $[m,1]$, respectively.
This gives the following model for the strain field $\bar{\boldsymbol{\epsilon}}(\mathbf{x})$ and the stress-free lattice spacing $d_{0}(\mathbf{x})$;
\begin{equation}\label{eq:basis_functions}
    \begin{split}
        &\bar{\boldsymbol\epsilon}_*(\mathbf{x}) = \boldsymbol\phi_\epsilon \mathbf{w}_\varphi, \quad \phi_{\epsilon,j}(\mathbf{x}) = \mathcal{V}^\mathbf{x}\phi_{\varphi,j}(\mathbf{x}), \quad \phi_{\varphi,j} = \frac{1}{\sqrt{L_{\varphi x}L_{\varphi y}}}\sin(\lambda_{\varphi x,j}(x + L_{\varphi x}))\sin(\lambda_{\varphi y,j}(y + L_{\varphi y})),\\
        &d_{0*}(\mathbf{x}) = \boldsymbol\phi_{d_0}\mathbf{w}_{d_0}, \quad \phi_{d_0,k}(\mathbf{x}) = \frac{1}{\sqrt{L_{d_0 x}L_{d_0 y}}}\sin(\lambda_{d_0 x,k}(x + L_{d_0 x}))\sin(\lambda_{d_0 y,k}(y + L_{d_0 y})),
    \end{split}
\end{equation}
where the unknown coefficients are independently normally distributed to approximate our GP model; $w_{\varphi,j} \sim\mathcal{N}(0,S_\varphi(\boldsymbol\lambda_{\varphi,j}))$ and $w_{d_0,j} \sim\mathcal{N}(\mu_{d_0,j},S_{d_0}(\boldsymbol\lambda_{d_0,j}))$. 
Where the means $\mu_{d_0,j}$ are chosen so that the prior has the constant value $\bar{d}_0$.
In this work, basis functions and parameters corresponding to the $d_0$ field will be denoted by the subscript $d_0$ and the subscript $k$ will be used as an index.
Likewise, basis functions and parameters corresponding to the Airys stress function will be denoted by the subscript $d_0$ and the subscript $j$ will be used as an index.
The expanded expressions for $\boldsymbol\phi_\epsilon$ are given in Appendix~\ref{app:basis_functions_and_their_derivatives}.

Using the LRT \eqref{eq:LRT} we can write a model for a predicted measurement as a non-linear function of the unknown coefficients;
\begin{equation}\label{eq:predicted_measurements}
\begin{split}
    y_* &= \frac{1}{L}\int\limits_0^L \bar{\mathbf{n}}\left(\sum_j \sum_k \phi_{\epsilon, j}(\mathbf{p}+\hat{\mathbf{n}}s)w_{\varphi,j}\phi_{d_0,k}(\mathbf{p}+\hat{n}s)w_{d_0,k}\right) + \left(\sum_k \phi_{d_0,k}(\mathbf{p}+\hat{n}s)w_{d_0,k}\right) \,\mathrm{d}s \\
     &= g_y(\mathbf{w}_\varphi,\mathbf{w}_{d_0},\boldsymbol\eta),
\end{split}
\end{equation}
where we have restricted ourselves to a single measurement to simplify the notation. 
These integrals can be analytically evaluated and the equations are given in Appendix~\ref{app:basis_functions_and_their_derivatives}.
Predictions of the boundary tractions $\mathbf{y}_t$ at a boundary location $\mathbf{x}_b$ with surface normal $\mathbf{n}_\perp$ can be written as a linear function of the unknown coefficients;
\begin{equation}\label{eq:predicted_tractions}
\begin{split}
    y_{t*} &= \underbrace{\pmat{n_{\perp1} & n_{\perp2} & 0 \\ 0 & n_{\perp1} & n_{\perp2} }}_{\mathbf{T}}\boldsymbol\phi_\varphi(\mathbf{x})\mathbf{w}_\varphi= \mathbf{T}(\mathbf{n}_\perp)\boldsymbol\phi_\varphi(\mathbf{x}_b)\mathbf{w}_\varphi \\
     &= g_t(\mathbf{w},\mathbf{x}_b,\mathbf{n}_\perp).
\end{split}
\end{equation}
The coefficients $\mathbf{w}_\varphi$ and $\mathbf{w}_{d_0}$ are random variables; as such the predictions $\bar{\boldsymbol{\epsilon}}_*$, $y_*$, and $y_{t*}$ are also random variables. 
The problem now is to determine the distribution of the coefficients given a set of LRT and traction measurements.
This problem is now in a form allowing variational inference to be used to approximate a solution to the non-linear problem.


\subsection{Variational Inference} 
\label{sec:variational_inference}
Variational inference \citep{BleiKM:2017,jordan1999introduction} provides an approximation to the posterior distribution by assuming that it has a certain functional form that contain unknown parameters. These unknown parameters are found using optimization, where some distance measure is minimized. We will in this section provide the details enabling the use of variational inference in solving our problem. 

Given $n$ transmission measurements and $n_t$ traction measurements, such that a vector of all measurements is given by $\mathbf{Y} = [y_1,\dots,y_n, y_{t,1},\dots,y_{t,n_t}]^\Transp$, the problem can be written as having prior and likelihood 
\begin{equation}
\begin{split}
    p(\mathbf{w}) \sim \mathcal{N}(\boldsymbol{\mu},\boldsymbol\Sigma_p) \ \ \text{and} \ \ p(\mathbf{Y} | \mathbf{w}) \sim \mathcal{N}(\mathbf{Y}| g(\mathbf{w}),\boldsymbol\Sigma_n), \\
\end{split}
\end{equation}
where $\mathbf{w} = \pmat{\mathbf{w}_{d_0}^\Transp & \mathbf{w}_\varphi^\Transp}^\Transp$, $\boldsymbol{\mu}$ is a vector of all the prior means, and $\boldsymbol\Sigma_p$ is a matrix with the coefficients prior variance on the diagonals. Here, $g(\cdot)$ is the combined measurement model that expresses the measurement vector $\mathbf{Y}$ as a function of the coefficients. This function is constructed using both \eqref{eq:predicted_measurements} and \eqref{eq:predicted_tractions}. Finally, $\boldsymbol\Sigma_n = \text{diag}(\sigma_n^2 I_{n\times n},\sigma_t^2 I_{n_t\times n_t})$, where $\sigma_t^2$ is a small variance placed on the artificial traction measurements added for numerical reasons.

The non-linear measurement function $g(\cdot)$ makes the likelihood intractable as the prior and likelihood are no longer conjugate. Consequently, the posterior $p(\mathbf{w}|\mathbf{Y})$ is also intractable and so we find an approximate solution using variational inference \citep{jordan1999introduction}.
The idea is to approximate the true posterior by the Gaussian distribution $q(\mathbf{w})\sim\mathcal{N}(\hat{\mathbf{w}}, \mathbf{C})$, and find the mean $\hat{\mathbf{w}}$ and covariance $\mathbf{C}$ for this distribution that maximise the Free Energy $\mathcal{F}$. 
The Free Energy places a lower bound on the log marginal likelihood and hence provides a measure of how well our posterior fits the data;
\begin{equation}\label{eq:free_energy}
    \log p(\mathbf{Y}) \geq \mathbb{E}\left[\log p(\mathbf{Y}|\mathbf{W})\right] - KL\left[q(\mathbf{w})||p(\mathbf{w}|\mathbf{Y})\right] = \mathcal{F}
\end{equation}
where, in this case, $\mathbb{E}[\cdot]$ is the expected value with respect to the approximate posterior $q(\mathbf{w})$ and $KL[\cdot]$ is the Kullback Leibler divergence which provides a measure of difference between the approximate posterior and the true posterior.
These terms can be evaluated as \citep{steinberg2014extended};
\begin{equation}
    \begin{split}
        \mathbb{E}\left[\log p(\mathbf{Y}|\mathbf{W})\right] &=\frac{1}{2}\left[N\log2\pi + \log|\boldsymbol\Sigma_n|+(\mathbf{Y}-\mathbb{E}\left[g(\mathbf{w})\right])^\Transp\boldsymbol\Sigma_n^{-1}(\mathbf{Y}-\mathbb{E}\left[g(\mathbf{w})\right]) \right],\\
        KL\left[q(\mathbf{w})||p(\mathbf{w}|\mathbf{Y})\right] &= \frac{1}{2}\left[\text{tr}(\boldsymbol\Sigma_p^{-1}\mathbf{C})+\left(\boldsymbol{\mu}-\hat{\mathbf{w}}\right)^\Transp\boldsymbol\Sigma_p^{-1}\left(\boldsymbol{\mu}-\hat{\mathbf{w}}\right)-\log|\mathbf{C}| + \log|\boldsymbol\Sigma_p|-N\right],
    \end{split}
\end{equation}
where $N = n+n_t$. Here, the expectation of the non-linear function $\mathbb{E}\left[g(\mathbf{w})\right]$ is intractable  \citep{steinberg2014extended} and so the expected maximum is used $\hat{\mathbf{Y}} = g(\hat{\mathbf{w}})$;
\small
\begin{equation}
\begin{split}
    \mathcal{F} &\approx -\frac{1}{2}\big[N\log2\pi +\log|\boldsymbol\Sigma_n| - \log|\mathbf{C}| + \log|\boldsymbol\Sigma_p|+(\mathbf{Y}-g(\hat{\mathbf{w}}))^\Transp\boldsymbol\Sigma_n^{-1}(\mathbf{Y}-g(\hat{\mathbf{w}})) + \left(\boldsymbol{\mu}-\hat{\mathbf{w}}\right)^\Transp\boldsymbol\Sigma_p^{-1}\left(\boldsymbol{\mu}-\hat{\mathbf{w}}\right)\big]
\end{split}
\end{equation}
\normalsize

The optimal posterior mean is chosen to maximise the Free Energy.
To perform this optimisation a modified Newton's method is used where the step direction is $\mathbf{q} = -\mathbf{H}^{-1}\mathbf{g}$ and we can calculate the gradient, $\mathbf{g}$, and Hessian, $\mathbf{H}$, of the cost as
\begin{equation}\label{eq:grad_hessian}
    \begin{split}
        \mathbf{g} &= \mathbf{J}^\Transp\boldsymbol\Sigma_n^{-1}(\mathbf{Y}-g(\hat{\mathbf{w}})) - \boldsymbol\Sigma_p^{-1}\hat{\mathbf{w}} \\
        \mathbf{H} &= -\mathbf{J}^\Transp\boldsymbol\Sigma_n^{-1}\mathbf{J} + \frac{\partial \mathbf{J}^\Transp}{\partial \mathbf{w}}\Sigma_n^{-1}(\mathbf{Y}-g(\hat{\mathbf{w}})) - \boldsymbol\Sigma_p^{-1} 
    \end{split}
\end{equation}
where $\mathbf{J} = \pmat{\frac{\partial \hat{\mathbf{Y}}}{\partial \hat{\mathbf{w}}}^\Transp & \frac{\partial \hat{\mathbf{Y}_t}}{\partial \hat{\mathbf{w}}}^\Transp}^\Transp $, and the derivatives and second derivatives are given in Appendix~\ref{app:measurement_model_derivatives}. 
At each iteration we update the coefficients according to
\begin{equation}\label{eq:update}
        \hat{\mathbf{w}}_{k+1} = (1-\alpha)\hat{\mathbf{w}}_k + \alpha \mathbf{q} + \alpha\boldsymbol{\mu},
\end{equation}
A backwards line search is used to ensure that $\mathcal{F}$ is increased in each iteration. 
Once the optimal posterior mean is found, the covariance can be found by setting $\frac{\partial \mathcal{F}}{\partial \mathbf{C}} = 0$ and linearising about $\hat{\mathbf{w}}$ \citep{steinberg2014extended}, giving;
\begin{equation}\label{eq:covariance}
    \mathbf{C} = \left[\boldsymbol\Sigma_p^{-1}+\mathbf{J}^\Transp \Sigma_n^{-1}\mathbf{J}\right]^{-1}.
\end{equation}
Pseudo-code for an algorithm to find approximate distribution of the coefficients $q(\mathbf{w})\sim\mathcal{N}(\hat{\mathbf{w}}, \mathbf{C})$ is given in Algorithm~\ref{alg:psudo_code}. 
Once the coefficients are found, estimates of the strain and $d_0$ fields can be estimated.
The approximate poster mean and variance for the strain field and stress-free lattice spacing can be computed as 
\begin{equation}
    \begin{split}
        \pmat{\hat{d}_0(\mathbf{x}) \\ \hat{\bar{\boldsymbol{\epsilon}}} } &= \pmat{\boldsymbol\phi_\epsilon(\mathbf{x}) & 0 \\ 0 &\boldsymbol\phi_\epsilon(\mathbf{x})}\hat{\mathbf{w}},  \\
        \hat{\Sigma}&= \pmat{\boldsymbol\phi_\epsilon(\mathbf{x}) & 0 \\ 0 &\boldsymbol\phi_\epsilon(\mathbf{x})}\mathbf{C} \pmat{\boldsymbol\phi_\epsilon(\mathbf{x}) & 0 \\ 0 &\boldsymbol\phi_\epsilon(\mathbf{x})}^\Transp,
    \end{split}
\end{equation}
where $\hat{\Sigma}$ is the joint covariance of the strain and $d_0$ estimates. Next, this method is demonstrated on a set of measurements simulated from a theoretical cantilever beam strain field and an artificial $d_0$ field.

\begin{algorithm}[htb]
      \caption{Variational inference algorithm for finding the coefficients $q(\mathbf{w})\sim\mathcal{N}(\hat{\mathbf{w}}, \mathbf{C})$. Requires the hyperparameters $\theta$, the specified number of basis functions $m_\varphi$ and $m_{d_0}$, the LRT measurement information $\{y_i, \boldsymbol{\eta}_i | \forall i = 1,\dots,n\}$ and the boundary traction information $\{y_{t,i}=0,\mathbf{x}_{b,i},\mathbf{n}_{\perp,i}|\forall i=1,\dots,n_t\}$.}
      \label{alg:psudo_code}
        \begin{algorithmic}[1]
            \Procedure{Find Coefficients}{}
            \State Compute the basis functions for the LRT measurements using Equation~\ref{eq:predicted_measurements}
            \State Compute the basis functions for the traction measurements using Equation~\ref{eq:predicted_tractions}
            \State Build prior variance $\boldsymbol\Sigma_p$
            \State Initialise the coefficients $\hat{\mathbf{w}}_1$
            \State set $k=1$
            \While{Stopping criteria not met} 
                \State Compute the gradient $\mathbf{g}$ and Hessian $\mathbf{H}$ linearised about $\hat{\mathbf{w}}_k$ according to Equation~\ref{eq:grad_hessian}
                \State Calculate $\hat{\mathbf{w}}_{k+1}$ using Equation~\ref{eq:update} and a backward line search
                \State $k = k+1$
            \EndWhile
            \State Calculate the covariance $\mathbf{C}$ according to Equation~\ref{eq:covariance}
          \State \textbf{return} $\hat{\mathbf{w}}_k$ and $\mathbf{C}$
          \EndProcedure
        \end{algorithmic}
    \end{algorithm}



\section{Simulation Results} 
\label{sec:results}
The method's ability to jointly reconstruct the strain field and a $d_0$ field is demonstrated using simulated measurements. 
Reconstructions from measurements simulated through two strain fields is shown; the Saint-Venant approximate strain field for a cantilver beam, and a Finite Element Analysis (FEA) strain field from an in-situ loaded \textsf{C}-shape.
Additionally, the consequences of ignoring the $d_0$ variation on the reconstruction are shown by using the linear measurement model and Gaussian process regression method presented by \citet{jidling2018probabilistic} with the addition of traction constraints as shown in \cite{hendriks2018traction}.
Matlab code to run both examples can be found on Github \citep{githubcode}.

\subsection{Cantilever Beam Example} 
\label{sub:cantilever_beam_eample}
\begin{figure}[!ht]
    \centering
    \includegraphics[width=0.4\linewidth]{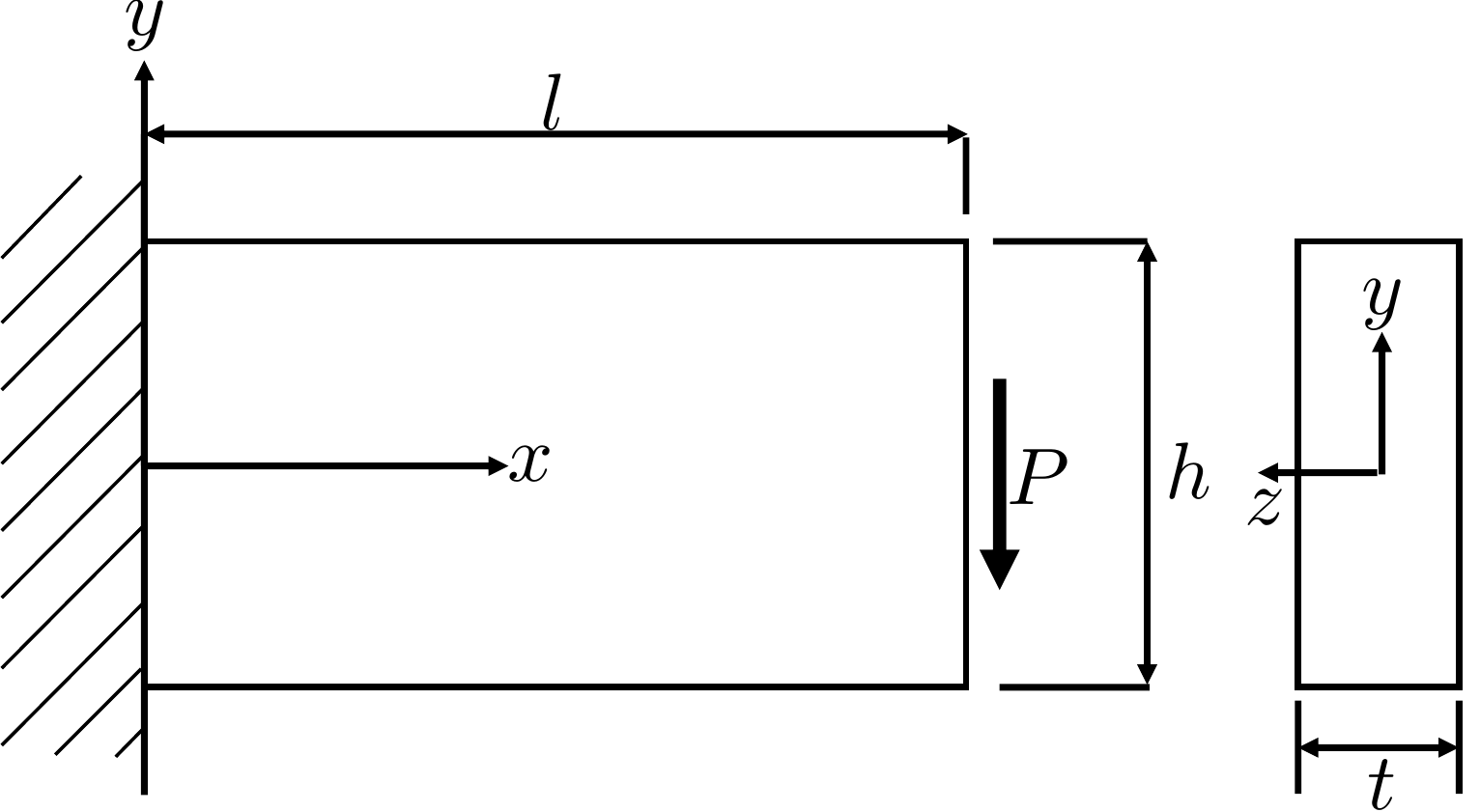}
    \caption{Cantilever beam geometry and coordinate system with $l = 20\si{mm}$, $h=10\si{mm}$, $t=5\si{mm}$, $E = 200\si{\giga\pascal}$, $P = 2\si{\kilo\newton}$, $\nu = 0.28$, and $I = \frac{th^3}{12}$.}
    \label{fig:CB_geom}
\end{figure}

The method is first demonstrated for the theoretical Saint-Venant cantilever beam as studied.
Assuming plane stress, the Saint-Venant approximation to the strain field is \citep{beer2010mechanics}:
\begin{equation}
    \boldsymbol{\mathcal{E}}(\mathbf{x}) = \begin{bmatrix}
        \frac{P}{EI}(L-x)y  \\
        -\frac{(1+\nu)P}{2EI}\left(\left(\frac{h}{2}\right)^2-y^2\right) \\
        -\frac{\nu P}{EI}(L-x)y
    \end{bmatrix},
\end{equation}
where the geometry is defined in Figure~\ref{fig:CB_geom}. A synthetic stress-free lattice spacing field is defined by
\begin{equation}\label{eq:d0_sim_field}
    d_0(\mathbf{x}) = c_0 \exp\left(-\frac{1}{2}(x-c_1)^2/c_2^2-\frac{1}{2}(y-c_3)^2/c_4^2\right) + c_5,
\end{equation}
with the parameters given by $\{c_0,c_1,c_2,c_3,c_4,c_5\} = \{0.0168,0,7.5\times10^{-3},7\times10^{-3},6\times10^{-3},4.056\}$. 
The maximum variation $c_0$ from a constant base value, $c_5$, was chosen to reflect the possible maximum relative variation due to martensitic phase change in $0.8\%$ carbon steel.

Measurements of the form \eqref{eq:LRT} were simulated for $30$ angles evenly spaced between $0^\circ$ and $180^\circ$, with $100$ measurements per angle, which is on the conservative side based on past experiments \citep{gregg2018tomographic,hendriks2017bragg}.
The simulated measurements were corrupted with zero-mean noise of standard deviation $\sigma_n = c_5\times10^{-4}$ which is equilvalent to $\num{1e-4}$ standard deviation in strain representing the typical experimental noise \citep{hendriks2017bragg,gregg2018tomographic} 

\begin{figure}[!ht]
    \centering
    \includegraphics[width=0.7\linewidth]{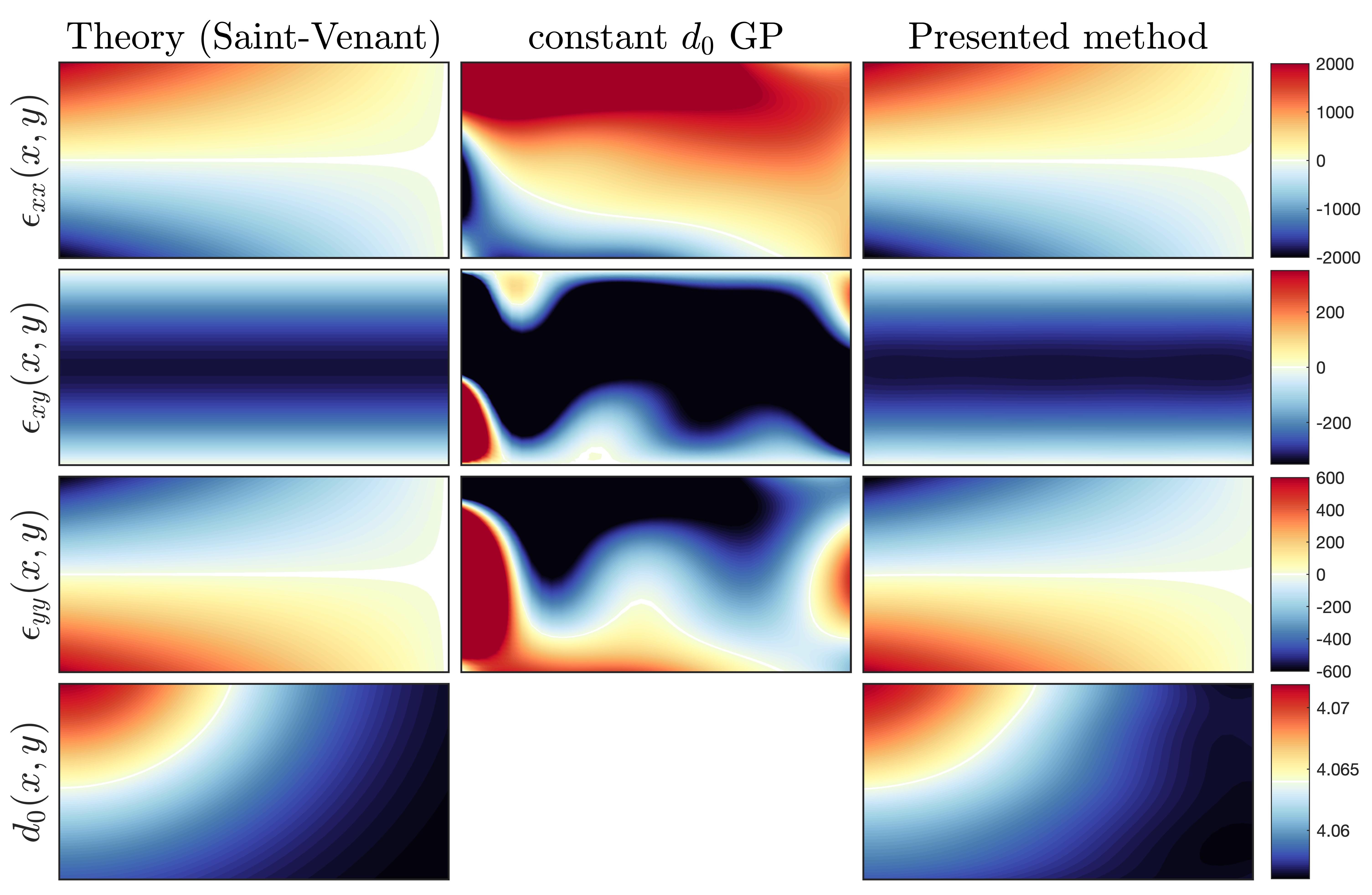}
    \caption{Simulation results for the cantilever beam strain field. The estimated strain field using the presented method is shown as well as the results of assuming a constant $d_0$ and applying standard GP regression. In the case of the presented method, the estimated $d_0$ field is also shown. Strain values are given in $\mu\text{Strain}$.}
    \label{fig:CB_results}
\end{figure}

Fifty zero-traction measurements were added along the top and bottom of the cantilever beam for both the presented method and the linear GP regression method.
Results are shown in Figure~\ref{fig:CB_results}. 
These results show that the presented method successfully reconstructs both the strain field and the $d_0$ field with a relative error\footnote{$\text{relative error} = \frac{\text{mean}\left(|\text{true}-\text{estimated}|\right)}{\text{max}\left(|\text{true}|\right)}$} of $0.0057$.
By contrast ignoring the presence of a $d_0$ variation and using a linear GP regression method yields a drastically degraded strain reconstruction with a relative error of $0.3067$.

\subsection{In-situ Loaded C-shape Sample} 
\label{sub:in_situ_loaded_c_shape_sample}
\begin{figure}[htb]
    \centering
    \includegraphics[width=0.3\linewidth]{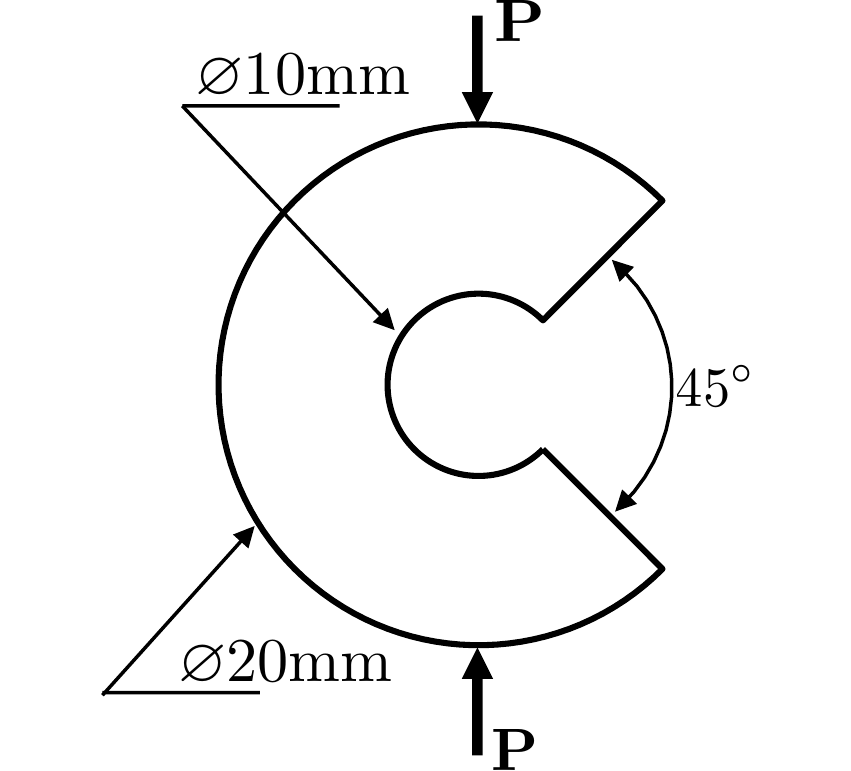}
    \caption{Geometry of the C-shape sample and in-situ loading $\mathbf{P}$. The sample has an outer diameter of $\SI{20}{\milli\metre}$ and an inner diameter of \SI{10}{\milli\metre} with a $45^\circ$ segment removed. The sample was defined to have $E = \SI{200}{\giga\pascal}$ and $\nu=0.28$.}
    \label{fig:C_geometry}
\end{figure}

\begin{figure}[htb]
    \centering
    \includegraphics[width=0.6\linewidth]{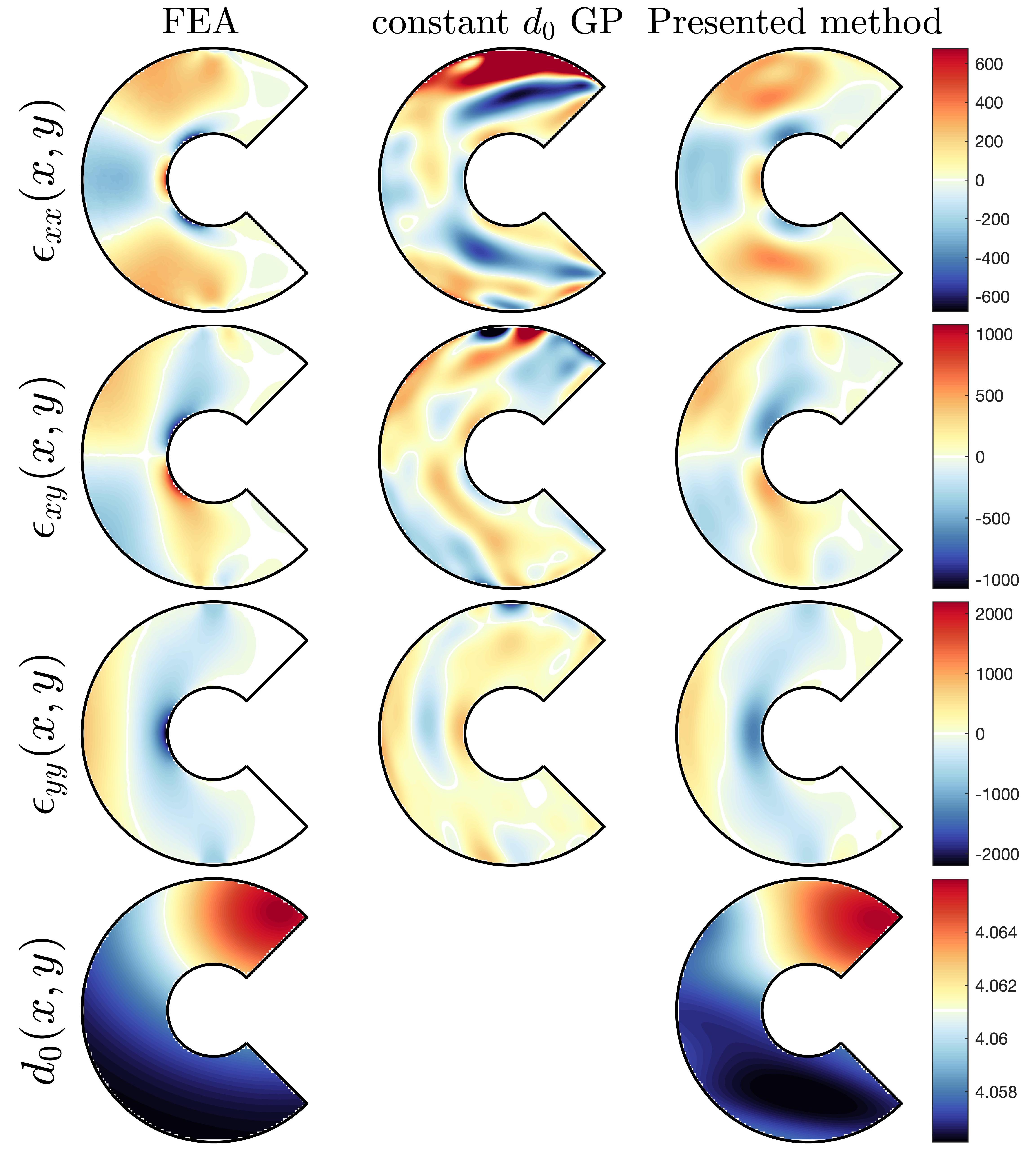}
    \caption{Simulation results for the FEA \textsf{C}-shape sample strain field. Strain values are given as $\mu\text{Strain}$. The estimated strain field using the presented method is shown as well as the results of assuming a constant $d_0$ and applying standard GP regression. In the case of the presented method, the estimated $d_0$ field is also shown.}
    \label{fig:C_shape_results}
\end{figure}

The method is now demonstrated on a more complex strain field given by FEA of a mild steel C-shape sample with geometry defined in Figure~\ref{fig:C_geometry}. The sample was subjected to a $\SI{7}{\kilo\newton}$ compressive load distributed over $5^\circ$ arcs and plane stress was assumed for the analysis. 
The resulting FEA strain field is shown in Figure~\ref{fig:C_shape_results}.
This sample and loading conditions correspond to the experimental setup used by \citet{hendriks2017bragg}. 

Measurements of the form \eqref{eq:LRT} were simulated through the this strain and a synthetic, smoothly changing $d_0$ field is again defined by Equation~\eqref{eq:d0_sim_field} with parameters given by $\{c_0,c_1,c_2,c_3,c_4,c_5\} = \{0.01,\num{5e-3},7.5\times10^{-3},7\times10^{-3},6\times10^{-3},4.056\}$. 
A total of 60 strain images were simulated with angles evenly spaced between $0^\circ$ and $180^\circ$, and 180 measurements per image. The simulated measurements were corrupted with zero-mean noise of standard deviation $\sigma_n = c_5\times10^{-4}$ which is equivalent to $\num{1e-4}$ standard deviation in strain representing the typical experimental noise.
A total of $131$ zero-traction measurements were added around the boundary of the \textsf{C}-shape excluding the regions within $10^\circ$ of the loading points.
Reconstruction from the LRT and traction measurements was performed using both the presented method and the linear GP regression method, and the results are shown in Figure~\ref{fig:C_shape_results}.

The presented method achieves a mean relative error of $0.023$ and it can be seen that the reconstruction has achieved the correct shape. Whereas assuming a constant $d_0$ value gives a mean relative error of $0.137$ and the resulting strain fields show incorrect concentrated peaks in the strain field and areas of tension and compression that are reversed. 
Despite this improvement there is still some observable difference. In particular, the presented method has a concentrated tensile region on the top left boundary of the \textsf{C}, and does not capture the very concentrated peaks in magnitude on the inside of the \textsf{C}.
These peak strains on the boundary are the hardest for the algorithm to reconstruct as they are poorly sampled by the LRT; i.e. they make up only a very small part of each line integral.
Additionally, some of this remaining error is due to systematic error in the simulation of the measurements.
Which are generated by numerically performing a line integral with each function evaluation being given by an interpolation of the FEA results.

\section{Additional Remarks} 
\label{sec:discussion}
\subsection{Sensitivity to the Traction Measurement Variance} 
\label{sub:necessity_of_tractions}
Boundary conditions given by unloaded surfaces are a natural inclusion as they are an artefact of the physical world. This information is included in the form of artificial measurements of zero traction, however it was found that a small variance needed to be placed on these measurements and the rate of convergence was impacted by the size of this variance.
Conceptually, this variance is analogous to a constraint tolerance for optimisation procedures.
Too large a variance (or not enough traction measurements) and the algorithm may fail to converge to the correct strain field. 
This indicates that the traction measurements are ensuring that the problem is observable, which is supported by the findings of \citet{hendriks2018robust} where the inclusion of traction measurements allowed a constant $d_0$ value to be found as a hyperparameter.
Conversely, too small a variance and the algorithm is unable to take optimisation steps of significant size, resulting in a large number of iterations to converge.
Methods for optimally choosing this variance is an avenue for future research.

Despite this, it was found that the algorithm worked well over a reasonable range of traction variances.
Typically the standard deviation of the traction measurements could be set two orders of magnitude smaller than the measurement standard deviation or in the range of $1\times10^{-5}$ to $1\times10^{-7}$.

\subsection{Hyperparameter Optimisation} 
\label{sub:hyperparameter_optimisation}
The hyperparameters $\theta = \{l_x,l_y,\sigma_f\}$ can be found by performing an optimisation using $\mathcal{F}$ as the objective function. However, the gradients of $\mathcal{F}$ with respect to the GP hyperparameters are not trivial and so \citet{steinberg2014extended} suggests that gradient free optimisation approaches could be used.
In this work, both Bayesian optimisation \citep{mockus2012bayesian} and the Nelda-Mead method \citep{nelder1965simplex} were found to work; with the Nelda-Mead method requiring less computation time.

\subsection{Computational cost} 
\label{sub:computational_cost}
The majority of the computational burden comes from building the building the forward model of the LRT measurements. 
In particular, all the combinations of the basis function of the stress-free lattice spacing and the strain field need to be computed for each measurement. The resulting matrix has $nm_{d_0}m_\varphi$ elements. 
However, as this matrix only needs to be computed once, it is still feasible to solve even with a large number of basis functions.


\section{Conclusion} 
\label{sec:conclusion}
This paper considers an extension of the strain tomography problem where the stress-free lattice parameter is a known constant, to the more general case where it is unknown and varies throughout the sample.
A method for the joint reconstruction of a strain field and a varying stress-free lattice parameter from a set of neutron transmission strain images has been presented. 
This method extends the Gaussian process based approach previously used for strain tomography to the subsequently non-linear problem, and ensures that the estimated strain fields satisfy equilibrium and can include knowledge of boundary conditions.
This was achieved by reformulating the problem in terms of basis functions and unknown coefficients.
Variational inference was then employed to find estimates of the coefficients.

The method was tested on a set of simulated data, and importantly, these results demonstrate that it is possible to perform this joint reconstruction.
Further, the results obtained by ignoring variations in $d_0$ and applying the linear GP regression method are provided and show that this assumption, if incorrect, severely degrades the accuracy of reconstruction.

Future work will involve planning an experiment to acquire a data set on which to further evaluate the methods performance.


\section*{Acknowledgements} 
\label{sec:acknowledgements}
This work is supported by the Australian Research Council through the Discovery Project scheme (ARCDP170102324), as well as the Swedish Foundation for Strategic Research (SSF) via the project \emph{ASSEMBLE} (contract number: RIT15-0012), and by the Swedish Research Council via the projects \emph{Learning flexible models for nonlinear dynamics} (contract number: 2017-03807) and \emph{NewLEADS - New Directions in Learning Dynamical Systems} (contract number: 621-2016-06079).


\begin{appendix}
\section{Basis Functions} 
\label{app:basis_functions_and_their_derivatives}
The basis functions for the strain field were defined in \eqref{eq:basis_functions} as
\begin{equation}
    \phi_{\epsilon, j} = \begin{bmatrix}
        \frac{\partial^2}{\partial y^2} - \nu\frac{\partial^2}{\partial x^2} \\
        -(1 + \nu)\frac{\partial^2}{\partial x \partial y} \\
        \frac{\partial^2}{\partial x^2} - \nu\frac{\partial^2}{\partial y^2}
    \end{bmatrix}\phi_{\varphi,j}.
\end{equation}
As such, the components of $\phi_{\epsilon,j}$ can be built from
\begin{equation}
\begin{split}
    \frac{\partial^2}{\partial x^2}\phi_{\varphi,j} &= \frac{-\lambda_{\varphi x,j}^2}{\sqrt{L_{\varphi x}L_{\varphi y}}}\sin(\lambda_{\varphi x,j}(x + L_{\varphi x}))\sin(\lambda_{\varphi y,j}(y + L_{\varphi y})), \\
    \frac{\partial^2}{\partial y^2}\phi_{\varphi,j} &= \frac{-\lambda_{\varphi y,j}^2}{\sqrt{L_{\varphi x}L_{\varphi y}}}\sin(\lambda_{\varphi x,j}(x + L_{\varphi x}))\sin(\lambda_{\varphi y,j}(y + L_{\varphi y})), \\
    \frac{\partial^2}{\partial x \partial y}\phi_{\varphi,j} &= \frac{\lambda_{\varphi x,j}\lambda_{\varphi y,j}}{\sqrt{L_{\varphi x}L_{\varphi y}}}\cos(\lambda_{\varphi x,j}(x + L_{\varphi x}))\cos(\lambda_{\varphi y,j}(y + L_{\varphi y})).
\end{split}
\end{equation}

A predicted LRT measurement was defined by \eqref{eq:predicted_measurements} as 
\begin{equation}
    y_*= \frac{1}{L}\int\limits_0^L \bar{\mathbf{n}}\left(\sum_j \sum_k \phi_{\epsilon, j}(\mathbf{p}+\hat{\mathbf{n}}s)w_{\varphi,j}\phi_{d_0,k}(\mathbf{p}+\hat{n}s)w_{d_0,k}\right) + \left(\sum_k \phi_{d_0,k}(\mathbf{p}+\hat{n}s)w_{d_0,k}\right) \,\mathrm{d}s
\end{equation}
where for clarity we restrict ourselves to a single measurement. 
Therefore, we need the components
\begin{equation}
    \begin{split}
        &\int\limits_0^L \phi_{d_0,k}(\mathbf{p}+\hat{\mathbf{n}}s)\,\mathrm{d}s , \\
        &\int\limits_0^L \phi_{d_0,k}(\mathbf{p}+\hat{\mathbf{n}}s)\left(\frac{\partial^2 }{\partial x^2}\phi_{\varphi,j}(\mathbf{p+\hat{\mathbf{n}}}s)\right)\,\mathrm{d}s , \\
        &\int\limits_0^L \phi_{d_0,k}(\mathbf{p}+\hat{\mathbf{n}}s)\left(\frac{\partial^2 }{\partial y^2}\phi_{\varphi,j}(\mathbf{p+\hat{\mathbf{n}}}s)\right)\,\mathrm{d}s , \\
        &\int\limits_0^L \phi_{d_0,k}(\mathbf{p}+\hat{\mathbf{n}}s)\left(\frac{\partial^2 }{\partial x \partial x}\phi_{\varphi,j}(\mathbf{p+\hat{\mathbf{n}}}s)\right)\,\mathrm{d}s , \\
    \end{split}
\end{equation}
To make the expressions briefer, we introduce the notation
\begin{equation}
\begin{split}
    \alpha_{\varphi x} &= \lambda_{\varphi x,j}(x_0 + \hat{n}_1s + L_{\varphi x}), \quad \alpha_{\varphi y} = \lambda_{\varphi y,j}(y_0 + \hat{n}_2s + L_{\varphi y}), \\
    \alpha_{d_0 x} &= \lambda_{d_0 x,k}(x_0 + \hat{n}_1s + L_{d_0 x}), \quad \alpha_{d_0 y} = \lambda_{d_0 y,k}(y_0 + \hat{n}_2s + L_{d_0 y}). \\
\end{split}
\end{equation}
Giving
\begin{subequations}\label{eq:meas_basis_funcs}
\begin{equation}
\begin{split}
    \zeta_k = &\int\limits_0^L \phi_{d_0,k}(\mathbf{p}+\hat{\mathbf{n}}s)\,\mathrm{d}s \\
     = &\frac{1}{\sqrt{L_{d_0 x}L_{d_0 y}}}\left(\frac{\sin(\alpha_{d_0 x} - \alpha_{d_0 y})}{2(\hat{n}_1\lambda_{d_0 x,k}-\hat{n}_2\lambda_{d_0 y,k})} - \frac{\sin(\alpha_{d_0 x} + \alpha_{d_0 y})}{2(\hat{n}_1\lambda_{d_0 x,k}+\hat{n}_2\lambda_{d_0 y,k})}\right)\Bigg|_{s=0}^{s=L}, \\
    \psi_{1,kj} = &\int\limits_0^L \phi_{d_0,k}(\mathbf{p}+\hat{\mathbf{n}}s)\left(\frac{\partial^2 }{\partial x^2}\phi_{\varphi,j}(\mathbf{p+\hat{\mathbf{n}}}s)\right)\,\mathrm{d}s \\
    = &\frac{-\lambda_{\varphi x,j}^2}{\sqrt{L_{\varphi x}L_{\varphi y}L_{d_0 x}L_{d_0 y}}}\left(-\Gamma_1-\Gamma_2+\Gamma_3+\Gamma_4-\Gamma_5+\Gamma_6-\Gamma_7+\Gamma_8\right)\Bigg|_{s=0}^{s=L},\\
    \psi_{2,kj} = &\int\limits_0^L \phi_{d_0,k}(\mathbf{p}+\hat{\mathbf{n}}s)\left(\frac{\partial^2 }{\partial y^2}\phi_{\varphi,j}(\mathbf{p+\hat{\mathbf{n}}}s)\right)\,\mathrm{d}s \\
     = &\frac{-\lambda_{\varphi y,j}^2}{\sqrt{L_{\varphi x}L_{\varphi y}L_{d_0 x}L_{d_0 y}}}\left(-\Gamma_1-\Gamma_2+\Gamma_3+\Gamma_4-\Gamma_5+\Gamma_6-\Gamma_7+\Gamma_8\right)\Bigg|_{s=0}^{s=L},\\
    \psi_{3,kj} = &\int\limits_0^L \phi_{d_0,k}(\mathbf{p}+\hat{\mathbf{n}}s)\left(\frac{\partial^2 }{\partial x \partial y}\phi_{\varphi,j}(\mathbf{p+\hat{\mathbf{n}}}s)\right)\,\mathrm{d}s \\
    = &\frac{\lambda_{\varphi x,j}\lambda_{\varphi x,j}}{\sqrt{L_{\varphi x}L_{\varphi y}L_{d_0 x}L_{d_0 y}}}\left(-\Gamma_1+\Gamma_2-\Gamma_3+\Gamma_4-\Gamma_5+\Gamma_6+\Gamma_7-\Gamma_8\right)\Bigg|_{s=0}^{s=L},\\
\end{split}
\end{equation}
where
\begin{equation}
\begin{split}
    \Gamma_1 &= \gd{-}{+}{+}, \\
    \Gamma_2 &= \gd{-}{-}{-}, \\
    \Gamma_3 &= \gd{-}{+}{-}, \\
    \Gamma_4 &= \gd{+}{-}{-}, \\
    \Gamma_5 &= \gd{+}{+}{-}, \\
    \Gamma_6 &= \gd{-}{-}{+}, \\
    \Gamma_7 &= \gd{+}{-}{+}, \\
    \Gamma_8 &= \gd{+}{+}{+}.
\end{split}
\end{equation}
Returning to the measurement model in \eqref{eq:predicted_measurements}, we can now write
\begin{equation}
    y_* = \frac{1}{L}\left(\sum_k\sum_j\bar{\mathbf{n}}w_{d_0,k}w_{\varphi j}\pmat{\psi_{2,kj}-\nu\psi_{1,kj}\\ -(1+\nu)\psi_{3,kj} \\ \psi_{1,kj} - \nu\psi_{2,kj}}_{s=0}^{s=L} + \sum_k w_{d_0, k}\zeta_k\big|_{s=0}^{s=L} \right).
\end{equation}
\end{subequations}

\section{Measurement Model Derivatives} 
\label{app:measurement_model_derivatives}
Here we give the derivates of the measurement model $g(\cdot)$ about the current $\hat{\mathbf{w}}$. 
The measurement model is a concatenation of equations \eqref{eq:predicted_measurements} and \eqref{eq:predicted_tractions}, and so we require the derivatives of the predicted LRT measurement, $\hat{y}$, and  the predicted traction $\hat{y}_t$. 
For clarity we restrict the following to a single $\hat{y}$ and $\hat{y}_t$.
The first derivatives are given by
\begin{equation}
\begin{split}
    &\frac{\partial \hat{y}}{\partial w_{d_0,k}} = \frac{1}{L}\left(\sum_j\bar{\mathbf{n}}w_{\varphi *, j}\pmat{\psi_{2,kj}-\nu\psi_{1,kj}\\ -(1+\nu)\psi_{3,kj} \\ \psi_{1,kj} - \nu\psi_{2,kj}}_{s=0}^{s=L} + \zeta_k\big|_{s=0}^{s=L} \right), \\
    &\frac{\partial \hat{y}}{\partial w_{\varphi,k}} = \frac{1}{L}\left(\sum_k\bar{\mathbf{n}}w_{d_0 *,k}\pmat{\psi_{2,kj}-\nu\psi_{1,kj}\\ -(1+\nu)\psi_{3,kj} \\ \psi_{1,kj} - \nu\psi_{2,kj}}_{s=0}^{s=L}\right), \\
    &\frac{\partial\hat{y}_t}{\partial \mathbf{w}_{d_0}} = 0, \\ 
    &\frac{\partial\hat{y}_t}{\partial \mathbf{w}_\varphi} = \mathbf{T}\boldsymbol\phi_\varphi(\mathbf{x}).
\end{split}
\end{equation}
The second derivates are
\begin{equation}
\begin{split}
    \frac{\partial^2 \hat{\mathbf{y}}_t}{\partial \mathbf{w}^2} &= 0, \\
    \frac{\partial^2 \hat{y}}{\partial \mathbf{w}^2}  &= \pmat{0 & \frac{\partial^2 \hat{y}}{\partial \mathbf{w}_{d_0} \partial \mathbf{w}_{\varphi}} \\ \left(\frac{\partial^2 \hat{y}}{\partial \mathbf{w}_{d_0} \partial \mathbf{w}_{\varphi}} \right)^\Transp & 0},
\end{split}
\end{equation}
where
\begin{equation}
    \left[\frac{\partial^2 \hat{y}}{\partial \mathbf{w}_{d_0} \partial \mathbf{w}_{\varphi}}\right]_{kj} = \bar{\mathbf{n}}\pmat{\psi_{2,kj}-\nu\psi_{1,kj}\\ -(1+\nu)\psi_{3,kj} \\ \psi_{1,kj} - \nu\psi_{2,kj}}_{s=0}^{s=L}.
\end{equation}
Explicit formulation of the second derivatives allows the cost functions curvature to be taken into account in the optimisation procedure, greatly improving the rate of convergence.


\end{appendix}


\bibliography{References}

\end{document}